# The Shapes of Galaxy Clusters

Pascal A.M. de Theije, Peter Katgert & Eelco van Kampen

*Sterrewacht Leiden, P.O.Box 9513, 2300 RA Leiden*

## ABSTRACT

We have reanalyzed a data set of 99 low redshift ($z < 0.1$) Abell clusters studied previously by Rhee, van Haarlem & Katgert (1989), and determined their shapes. For this, three different measures are used: two of which were originally used by Rhee et al., and one of which was used by Plionis, Barrow & Frenk (1991) in their investigation of clusters in the Lick catalogue. We use Monte-Carlo simulations of clusters to investigate the errors in the methods. For low ellipticity, all methods overestimate the cluster elongation, whereas the opposite is true for a highly flattened system. Also background galaxies and shot noise have a rather large influence on the measured quantities.

The corrected distribution of cluster ellipticities shows a peak at $\epsilon \sim 0.4$ and extends to $\epsilon \sim 0.8$, consistent with results of some previous studies. However, the present study uses more than twice the number of clusters as the earlier studies, and is self-consistent. That is, with the corrected distribution over projected cluster shapes we can reconstruct the observed distribution over projected cluster shapes and the observed relation between the number of galaxies in a cluster and its ellipticity. To achieve this, we have to assume that there is an anti-correlation between the true (projected) ellipticity of a cluster and its number of galaxies. It is not necessary to assume that the ellipticity of a cluster increases when one only includes the brighter galaxies (as suggested by Binney 1977).

Using a redshift-independent richness criterion of Vink & Katgert (1994), it is shown that the richer clusters are intrinsically more nearly spherical than the poorer ones. Furthermore, the corrected distribution of cluster shapes is found to be more consistent with a population that consists of purely prolate clusters than with a purely oblate population.

We compare the corrected true distribution of (projected) ellipticities with pre-dictions from N-body simulations. For this, we use a catalogue of 75 N-body simulated clusters (van Kampen 1994) which assume a CDM spectrum with $\Omega = 1.0$. The simulations include a recipe for galaxy formation and merging. The model clusters are expected to be a representative sample of all real clusters. They show a good resemblance with the data both in radial profile and number of galaxies. 'Observing' these simulated clusters in exactly the same way as the real clusters produces an ellipticity distribution that extends to much higher $\epsilon$ and that has too few nearly spherical clusters. Preliminary results of simulations of the formation of clusters in an $\Omega = 0.2$ universe suggest that, on average, clusters are more nearly spherical in this case, as is expected on theoretical grounds. This shows that the elongations of clusters can provide a useful constraint on the value of $\Omega$.

# 1   INTRODUCTION

Since clusters of galaxies have only just collapsed or are still collapsing (e.g. Regös & Geller 1989; van Haarlem 1992; Zabludoff & Franx 1993), their shapes will probably contain clues about the formation of large-scale structure and about the initial density field out of which structure in the Universe is believed to have formed by gravitational collapse.

Most clusters, like elliptical galaxies, are non-spherical, and their shape is (as is the case for ellipticals) not due to rotation (Rood et al. 1972; Gregory & Tifft 1976; Dressler 1981). The perturbations that gave rise to the formation of galaxy clusters are likely to have been aspherical initially (Barrow & Silk 1981; Peacock & Heavens 1985; Bardeen et al. 1986). The asphericities are then amplified during gravitational collapse as the collapse proceeds more rapidly along the short axis (Lin, Mestel & Shu 1965; Icke 1973; Barrow & Silk 1981). In cases where there are good measurements, the elongation seems to be due to a velocity anisotropy of the galaxies (Aarseth & Binney 1978), as in elliptical galaxies. Binney & Silk (1979) claim that the elongation of clusters originates in the tidal distortion by neighboring protoclusters. Recently, Salvador-Solé & Solanes (1993) also suggested this by developing a simple model, with which they can explain the elongations of clusters. They point out that the main tidal distortion on a cluster is by the nearest neighboring cluster having more than 45 galaxies. Their model also explains the (weak) alignment between neighboring clusters (Oort 1983; Binggeli 1982; Rhee & Katgert 1987; Plionis 1993) and the strong alignment between clusters and their first ranked galaxy (Sastry 1968; Carter & Metcalfe 1980; Dressler 1981; Binggeli 1982; Rhee & Katgert 1987; Tucker & Peterson 1988; van Kampen & Rhee 1990; Lambas et al. 1990; West 1994). Recently, Bertschinger & Jain (1993) and van de Weygaert & Babul (1994) claimed that shear fields probably have a large influence on the shapes of clusters. This shear can be due to the clusters own asphericity or can be induced by the surrounding mass distribution. It can break up a single cluster into more distinct ones, or promote a smaller cluster to a larger one. Another possible reason for the large elongations of clusters is given by Zeldovich (1978), who claims that it is due to the fact that clusters are born in sheets of gas ("pancakes"). However, this would require structures to have formed via a top-down or fragmentation scenario. In such a scenario clusters fragment out of larger structures and will have a certain elongation in the direction of the sheets. Although this scenario can also explain the alignment between neighboring clusters, it is probably ruled out (Peebles 1993) because there are clear examples of galaxies that are assembled before the systems in which they are now found. Also, the existence of high-redshift quasars is a problem for the top-down scenario.

Some N-body simulations have shown that the shapes of dissipationless dark halos of clusters most likely will contain very little information about the fluctuation spectrum giving rise to the perturbations (West, Dekel & Oemler 1989; Salvador-Solé 1993). On the other hand Quinn, Salmon & Zurek (1986), Efstathiou et al. (1988) and Crone, Evrard & Richstone (1994) demonstrated that there is a link between the final cluster structure and the initial spectrum of density fluctuations. It is not clear how this is for the luminous part of the cluster, i.e. the galaxies (van Kampen 1994).

A cosmological parameter that certainly influences the distribution of galaxies in a cluster is the density parameter $\Omega$. Evrard (1993) and Evrard et al. (1993) have shown, by comparing N-body simulations with X-ray images, that clusters in a low-$\Omega$ universe are much more regular, spherically symmetric, and centrally condensed than clusters in an $\Omega = 1$ universe. This is because in a low-$\Omega$ universe, structures will collapse at a larger redshift than in a high-$\Omega$ universe (Maoz 1990) and thus have had more time to virialize and wipe out their asymmetries and substructures. Furthermore, the timescale in a low-$\Omega$ universe is (much) longer than in an Einstein-de Sitter universe (see e.g. Padmanabhan 1993, eq. [2.76]). Evrard (1993) and Evrard et al. (1993) conclude that the complex nature of real cluster X-ray morphologies is better matched by an Einstein-de Sitter model than by $\Omega = 0.2$ models, even if they include a cosmological constant in the latter.

However, the observational information on the distribution of cluster shapes is not unambiguous. While Rhee, van Haarlem & Katgert (1989; hereafter RHK) find that most clusters are nearly spherical, with a peak at $\epsilon \sim 0.15$, Carter & Metcalfe (1980) , Binggeli (1982) and Plionis, Barrow & Frenk (1991; hereafter PBF) find them to be much more elongated, with most clusters having $\epsilon \sim 0.5$ ($\epsilon$ denotes the projected ellipticity of a cluster, defined as one minus the projected axis ratio).

In this paper we try to remove the inconsistency between the different determinations of the cluster shapes by comparing the methods used by the various authors. We concentrate on the samples used by RHK and PBF and apply the method of PBF to the data of RHK.

The outline of the paper is as follows: in Section 2, the sample of clusters is described. In Section 3, we describe the methods used by RHK and PBF. Section 4 outlines the Monte-Carlo simulations, used to correct for the different systematic and random errors in the various methods. In Section 5 we describe the corrected results and investigate the relation between the elongation of clusters and their richness. In Section 6, we try to put constraints on the distribution over the intrinsic ellipticities of the clusters by assuming them to have a single triaxiality parameter. In Section 7 we use the N-body simulations of clusters, taken from a catalogue of model clusters by van Kampen (1994), with either $b = 1.0, \Omega = 0.2$ or $b = 2.0, \Omega = 1.0$. The shapes of the luminous part of the clusters in these models are compared with the data. In Section 8 we give a discussion and summarize the results.

A Hubble constant $H_0 = 100h$ km s$^{-1}$ Mpc$^{-1}$ is used throughout the paper.

## 2  THE DATA SETS

The cluster samples are fully described by the authors (RHK and PBF). For completeness, the main properties are reviewed here.

### 2.1  RHK sample

The RHK sample consists of 99 Abell clusters of Abell richness $R \geq 1$, redshift $z \leq 0.1$, $10^h \leq \alpha \leq 18^h$, $\delta < -25^o$ and $b \geq 30^o$ (the original sample consists of 107 clusters, but only for the 99 clusters in Figure 5 of RHK all data were available). The sample is complete up to a redshift $z = 0.08$. The galaxy positions and brightnesses were determined using the Leiden Astroscan automatic plate measuring machine, from glass copies of the red Palomar Sky Survey plates. The machine has been described by Swaans (1981). To identify the galaxies, the following stages were gone through: first, all objects down to a certain minimum threshold level in surface brightness were detected. Then stellar objects were separated from the non-stellar objects, using a separation criterion based on that devised by Kron (1980). The object position was then defined as the mean position of the five brightest pixels of the nine pixels surrounding the first object position. The galaxy catalogues have magnitude limits of $m_R$ between 18.0 and 19.0.

The cluster centers were roughly determined using the coordinates in $mm$ given by Sastry & Rood (1971). Then the brightest galaxy within $\sim 0.25 h^{-1}$ Mpc was found, and all galaxies which were detected within $1h^{-1}$ Mpc from this brightest galaxy, were selected; note that RHK used $h = 0.5$ in their paper. Later, Vink & Katgert (1994) redetermined the centers by defining a circular area around the old center and calculating the center of mass of the galaxies in this area, which then became the new center. This procedure was repeated until the center did not change anymore. Convergence was always achieved.

To estimate the background, 16 randomly chosen field regions were scanned in exactly the same way as the cluster regions. Based on the galaxy counts in these 16 regions, it was found that the average field contamination to $m_R \sim 19$ is $\sim 50\%$ within $1h^{-1}$ Mpc from the cluster center.

### 2.2 PBF sample

The sample used by PBF consists of $\sim 100$ clusters (with $|b| \geq 40^{\circ}$) in the Lick map, based on the Shane & Wirtanen galaxy catalogue (Shane & Wirtanen 1967; Seldner et al. 1977). This catalogue contains 810,000 galaxies with $m_b \leq 18.8$ binned in $10 \times 10$ arcmin$^2$ cells. PBF smoothed the counts using a discrete convolution with the weights used by Shectman (1985). They then generated three cluster catalogues, using three overdensity thresholds viz. $\sigma/<\sigma> = 3.6, 2.5$ and $1.8$, corresponding to 5.0, 3.5 and 2.5 galaxies per cell, respectively ($\sigma$ denotes the number of galaxies per cell). They based their discussion of cluster ellipticities mainly on the 3.6 catalogue. In order to qualify as a cluster, a surface density peak needed to occupy at least 5 cells that are above the threshold (such cells were called 'active cells'). The clusters in the 3.6 catalogue have redshifts up to $z \sim 0.15$, with a median value of $\sim 0.07$.

## 3  THE METHODS

In this section, we outline the methods used by RHK and PBF to determine the shape of a cluster.

### 3.1  Description

RHK used two different methods, one based on moments (from now on called the moments method, its result indicated by $\epsilon_1$), the other based on the inertia tensor (henceforth called the tensor method, its result denoted by $\epsilon_2$).

For the moments method, the various second order moments of the galaxy distribution are defined as

$$\mu_{ab} = \frac{1}{N_{gal}} \sum x^a y^b \ , \quad (a,b = 0,1,2; \quad a+b = 2) \ , \tag{1}$$

where the sum is over all galaxies, $x$ and $y$ are the coordinates of the galaxy relative to the cluster center, and $N_{gal}$ is the number of galaxies in the cluster. The cluster ellipticity is then given by

$$\epsilon_1 = 1 - \sqrt{\frac{1-e}{1+e}} \ , \tag{2}$$

where

$$e = \frac{\sqrt{(\mu_{20} - \mu_{02})^2 + 4\mu_{11}^2}}{\mu_{20} + \mu_{02}} \ . \tag{3}$$

For the tensor method, the inertia tensor is defined as

$$I_{ij} = \sum x_i x_j / r^2 \ , \quad (i,j = 1,2) \ , \tag{4}$$

where the sum is over all galaxies, $x_1 = x$, $x_2 = y$ and $r$ is the projected distance of a galaxy to the cluster center. The cluster ellipticity is then given by

$$\epsilon_2 = 1 - \Lambda_{min}/\Lambda_{max} \ , \tag{5}$$

where

$$\Lambda_{min,max} = \frac{1}{2}(I_{11} + I_{22}) \pm \frac{1}{2}\sqrt{(I_{11} + I_{22})^2 - 4(I_{22}I_{22} - I_{12}I_{21})} \tag{6}$$

Both methods used by RHK calculate $\epsilon$ within 1 $h^{-1}$ Mpc from the center. A few clusters are located near the edge of a plate and $\epsilon$ is then evaluated at a smaller radius.

For the PBF method, the inertia tensor is defined as

$$I_{ij} = \sum x_i x_j m \ , \quad (i,j = 1,2) \ , \tag{7}$$

where the sum is over all active cells having more than $\sigma = 3.6 < \sigma \geq 5.0$ galaxies, $x_1 = x$ and $x_2 = y$ are the coordinates of the center of the cell with respect to the cluster center, and $m$ is the number of galaxies in that cell (after smoothing). The cluster ellipticity is then defined according to eqs. (5) and (6), and will be denoted by $\epsilon_3$. Only clusters having at least five active cells are included. This causes a systematic selection against non-centrally concentrated clusters.

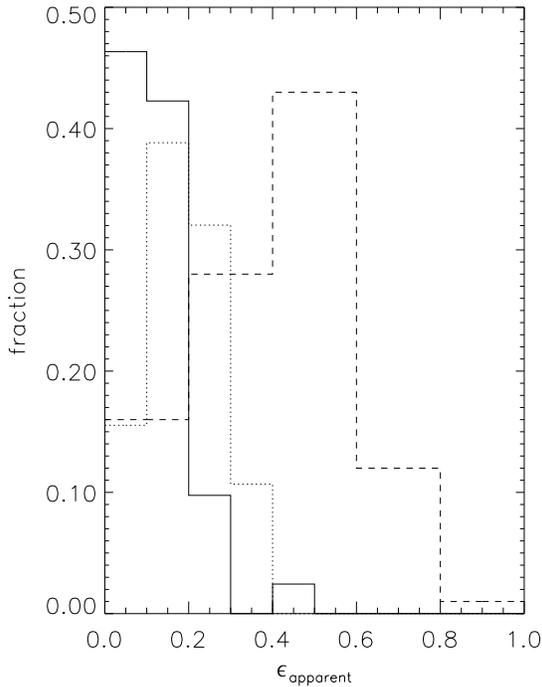

**Figure 1.** Original distributions over ellipticities of RHK and PBF. The solid line gives the distribution for the moments method of RHK, the dotted line denotes the tensor method of RHK (both derived from the RHK data), and the dashed line indicates the results from the tensor method of PBF and as derived from the PBF data.

### 3.2 Original results

The apparent (i.e., as originally measured by the authors) distributions over $\epsilon_1$, $\epsilon_2$ and $\epsilon_3$ are shown in Figure 1. The solid line corresponds to the moments method of RHK, the dotted line to the tensor method of RHK, and the dashed line to the tensor method of PBF. It is obvious from Figure 1 that the two methods of RHK give quite different results, especially at low $\epsilon_{apparent}$. The tensor method shows a lack of spherical clusters relative to the moments method. Since they are applied to exactly the same data, this clearly illustrates the effect of using different methods. The PBF method is even more different and finds many elongated clusters. However, this may partly be due to the different limiting magnitude of both samples (see Section 2). The figure shows the apparent inconsistency between different methods and maybe different data sets.

Because the RHK sample is almost fully available to us, from now on only this data set will be used and the presented results for the PBF method are obtained by applying this method to the RHK data set rather than to the PBF data set. Since the Lick map only contains objects with a blue apparent magnitude $m_b \leq 18.8$ and the RHK data set is based on red plates, one has to take a corresponding limit in the red band to compare both data sets. We have determined this limit by looking at the cumulative number of active cells (of the RHK sample) as a function of limiting magnitude (in the red), and compared this to the distribution over active cells of PBF (their Figure 7). The magnitude limit that gives the best fit is $m_R \sim 17.0$. This means, that if we apply a magnitude limit $m_R \sim 17.0$ to the RHK sample, we on average select the same galaxies as does the Lick catalogue. The conversion from blue to red magnitudes is in agreement with our photometric calibration (Vink & Katgert 1994) and with the average color $B - R \approx 1.8$, given by Carter & Metcalfe (1980) and Postman, Geller & Huchra (1988). Furthermore, we assume the cluster centers of the RHK sample to be randomly positioned with respect to the grid of cells that is needed to apply the PBF method. This may introduce some errors, because the measured ellipticity of a cluster is found to depend on the positioning of the center with respect to the cells. As mentioned earlier, a rather large fraction of the clusters are rejected because they do not have the 5 active cells required to include them in the calculations.

## 4 MONTE-CARLO SIMULATIONS

In this section we apply the methods described in the Section 3.1 to computer-generated (Monte-Carlo) clusters. In this way the relations between input ('true') and output ('apparent') ellipticity can be established. These relations are then used to statistically correct the distributions of measured ellipticities for systematic effects in the methods.

### 4.1 Generating synthetic clusters

For the model clusters we assume a triaxial density distribution according to a modified Hubble-profile, and stratified on similar, concentric ellipsoids,

$$\rho(m) = \rho_0[1 + m^2]^{-3/2} \; , \tag{8}$$

where

$$m^2 = \frac{x^2}{a^2} + \frac{y^2}{b^2} + \frac{z^2}{c^2} \; , \quad a \geq b \geq c \; , \tag{9}$$

(Sarazin 1986, and references therein; RHK). $\rho_0$ is the central galaxy density in the cluster. This profile, together with assumed values $b/a$ and $c/a$ of the intrinsic ellipticities, serves as the probability function from which the galaxies are drawn randomly. The projected surface density profile, associated with equation (8), is

$$\mu(\tilde{m}) = \mu_0[1 + \tilde{m}^2]^{-1} \; , \tag{10}$$

where $\mu_0 = 2a\rho_0$ and

$$\tilde{m}^2 = \frac{\tilde{x}^2}{\tilde{a}^2} + \frac{\tilde{y}^2}{\tilde{b}^2} \; , \tag{11}$$

with $\tilde{x}, \tilde{y}$ the projected coordinates on the sky, and $\tilde{a}$ the cluster core radius. A value $\tilde{a} \sim 0.25 h^{-1}$ Mpc is adopted for the core radius (Sastry & Rood 1971; Bahcall 1975; Dressler 1978; Binggeli 1982; Jones & Forman 1984; Sarazin 1986, and references therein; RHK). See, however, Beers & Tonry (1986), who claim that the core radius may be much smaller or even disappear if not the galaxy positions but e.g. cD galaxies are used to determine the cluster center.

Furthermore, we include a background in the simulations. The background galaxies are distributed at random over an area of $[-2, 2] \times [-2, 2] h^{-1}$ Mpc around the cluster center. Such a large area is chosen because the PBF-method is not limited to a fixed area, so one has to include enough background. To include the right number of background galaxies, the magnitude distribution of the field galaxies in the 16 sky fields of RHK is used together with an assumed magnitude limit. These field galaxy counts are in agreement with results of Binggeli (1982), Jones et al. (1991) and Metcalfe et al. (1991). Although the background galaxies will, in reality, not be distributed randomly over the area because galaxies are believed to lie in walls and filaments, and those that are near a cluster may experience the influence from that cluster, the random positioning is found to give a good approximation. In fact, the results depend only very slightly on the precise number of background galaxies.

Similar simulations are done for core radii of 0.1 and $0.5 h^{-1}$ Mpc. They give very similar results for the correction cubes, and so our results will be rather insensitive to the actual value of the adopted core radius.

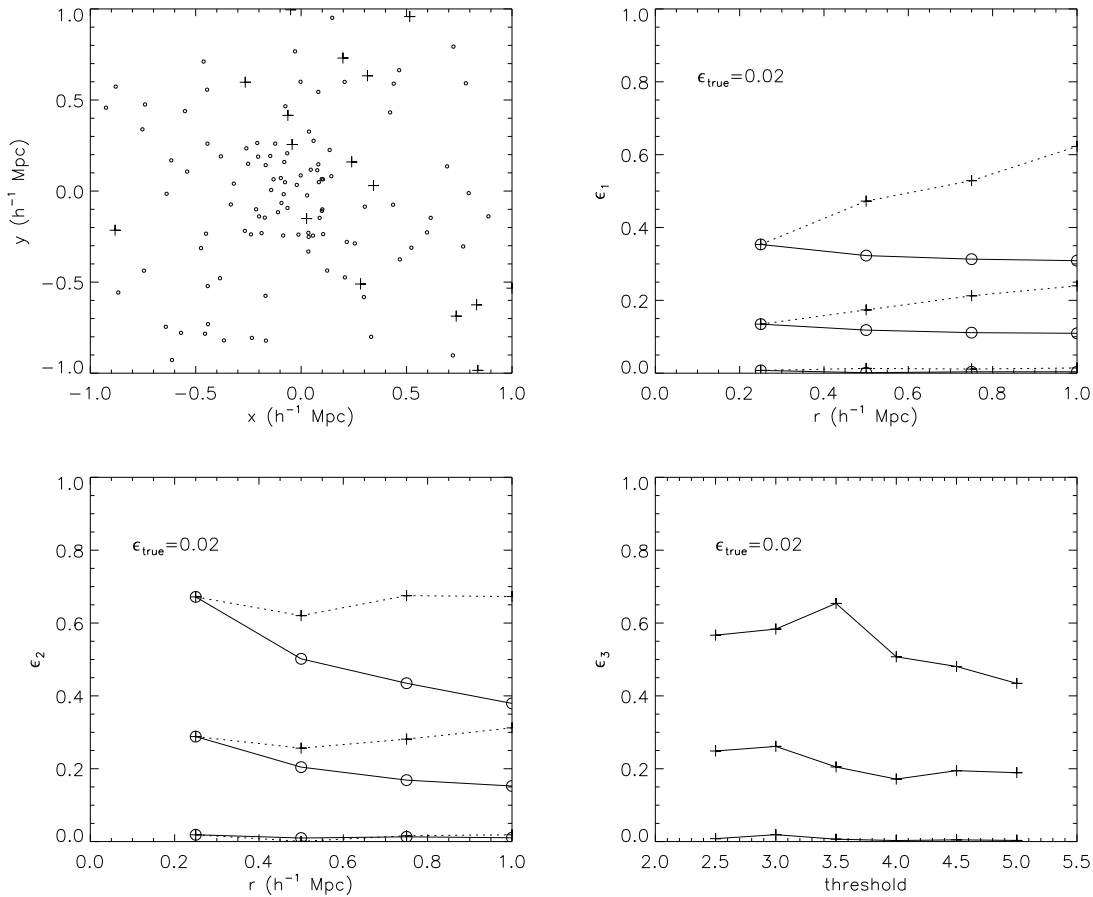

**Figure 2.** Example of a Monte-Carlo cluster. This cluster has a true projected ellipticity of 0.02, contains 150 member galaxies, and 60 background galaxies randomly distributed over the area of $[-2, 2] \times [-2, 2] h^{-1}$ Mpc around the cluster center. The upper lefthand plot shows the galaxy distribution of one of the realizations. Circles are cluster galaxies, crosses indicate background galaxies. The upper righthand plot shows the variation of $\epsilon_1$ with radius, as derived from 500 realizations. Circles and solid lines are cumulative values, whereas crosses and dotted lines are differential values. The middle lines indicate the average value of the ellipticity over the 500 realizations. The upper and lower lines denote the maximal and minimal value of $\epsilon_1$, respectively. The lower left plot shows the variation of $\epsilon_2$ with radius. The lower right plot shows the variation of $\epsilon_3$ with overdensity threshold.

### 4.2 Runs

The Monte-Carlo simulations are performed as follows: We construct clusters with two intrinsic axial ratios $b/a$ and $c/a$ and for given viewing angles one can then calculate the projected ellipticity $\tilde{b}/\tilde{a}$. For a fixed projected ellipticity $\epsilon_{true} = \tilde{b}/\tilde{a}$ the galaxy distribution on the sky will always be the same, independent of the intrinsic three-dimensional shape. This is because of the assumption of similar, concentric ellipsoids (e.g. Stark 1977). For each projected ellipticity, $\epsilon_{true} = 0.00, 0.05, 0.10, ..., 1.00$, 500 clusters are generated, according to the above constraints. An example of such a Monte-Carlo cluster is shown in Figure 2, together with the ellipticities as measured by the three methods. This cluster has a true projected ellipticity of 0.02, contains 150 member galaxies distributed according to equation (10), and 60 background galaxies randomly distributed over the area of $[-2,2] \times [-2,2]h^{-1}$ Mpc around the cluster center. The upper lefthand plot shows the galaxy distribution of one of the realizations. Circles are cluster galaxies, crosses indicate background galaxies. The upper righthand plot shows the variation of $\epsilon_1$ with radius. Circles and solid lines are cumulative values, crosses and dotted lines are differential values. The middle lines indicate the average value of the ellipticity over the 500 realizations. The upper and lower lines denote the maximal and minimal value of $\epsilon_1$, respectively. The lower lefthand plot shows the variation of $\epsilon_2$ with radius. The symbols and lines have the same meaning as in the upper righthand plot. The lower righthand plot shows the variation of $\epsilon_3$ with threshold. The middle line again denotes the mean, the upper and lower lines denote the maximal and minimal values of $\epsilon_3$, respectively. This figure shows some striking features. First, the apparent ellipticity changes with radius. This is best seen in the lower left plot. It is only caused by noise in the method because the cluster is constructed to have a constant $\epsilon_{true}$ with radius. Second, there is a large spread in apparent ellipticities, which decreases with radius. This is due to the smaller number of galaxies in the center. Third, the average measured $\epsilon_{apparent}$ is always larger than the true one (in this particular case of $\epsilon_{true} = 0.02$; but see later).

The above procedure is performed for clusters containing different numbers of member galaxies, $N_{gal}$, ranging from 50 to 1000. This generates a 'correction cube', which gives the distribution over $\epsilon_{apparent}$ for a range of $\epsilon_{true}$-values and a range of $N_{gal}$-values. With this cube we are able to correct statistically for the systematic errors and noise in the different methods. Figure 3 shows a slice through this cube, for clusters having 100 member galaxies and 60 background galaxies. The contours are equidistant and reflect the probability that a true ellipticity $\epsilon_{true}$ results in a measured ellipticity $\epsilon_i$ ($i = 1, 2, 3$).

### 4.3 Results of simulations

For small numbers of galaxies, the moments method of RHK gives $\epsilon_1 \sim 0.10$ for $\epsilon_{true} \leq 0.15$. For higher $\epsilon_{true}$, $\epsilon_1$ increases too, but is always well below the true value. At $\epsilon_{true} = 1.0$, $\epsilon_1 \sim 0.7$. The method performs better for larger numbers of galaxies, i.e., the measured curve approaches the theoretical curve and the distribution over $\epsilon_1$ gets narrower, indicating that the spread in the measured values decreases.

The results for the tensor method of RHK are similar to those for the moments method, except that the method is slightly less accurate for small $\epsilon_{true}$ (in absolute sense; the width of the distributions are similar), and more accurate for higher $\epsilon_{true}$. The dependence on the number of galaxies is similar.

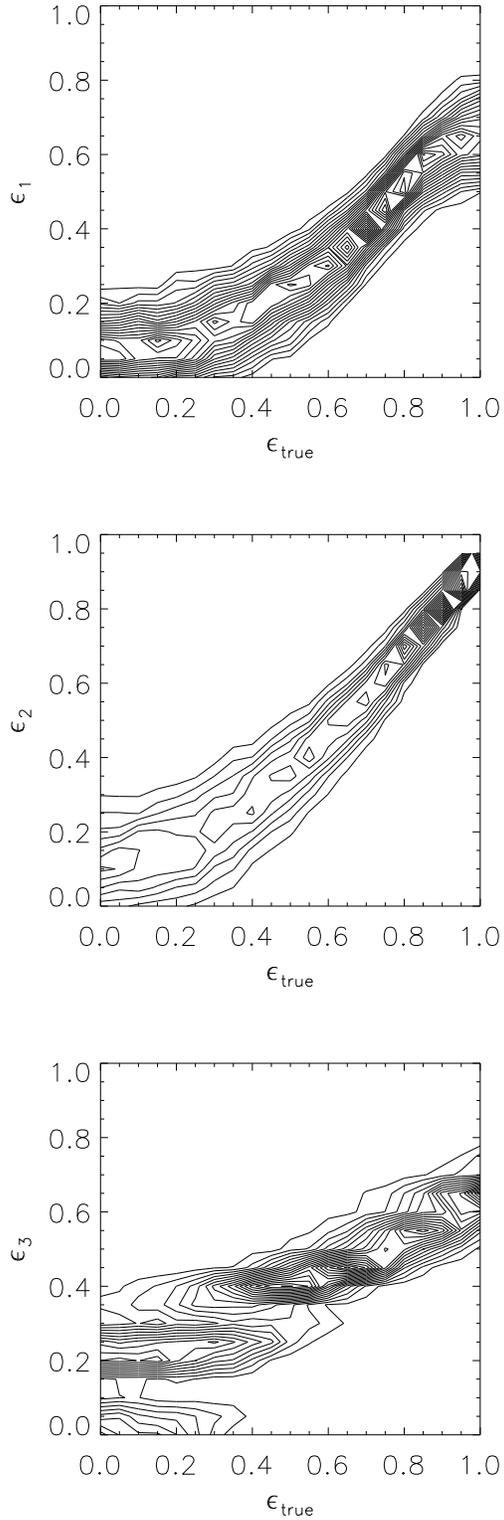

**Figure 3.** Relation between true and apparent ellipticity for clusters containing 100 member galaxies. A value of 60 background galaxies is adopted. The upper plot is for the moments method of RHK, the middle plot is for the tensor method of RHK and the lower plot is for the tensor method of PBF. The contours are at equidistant values.

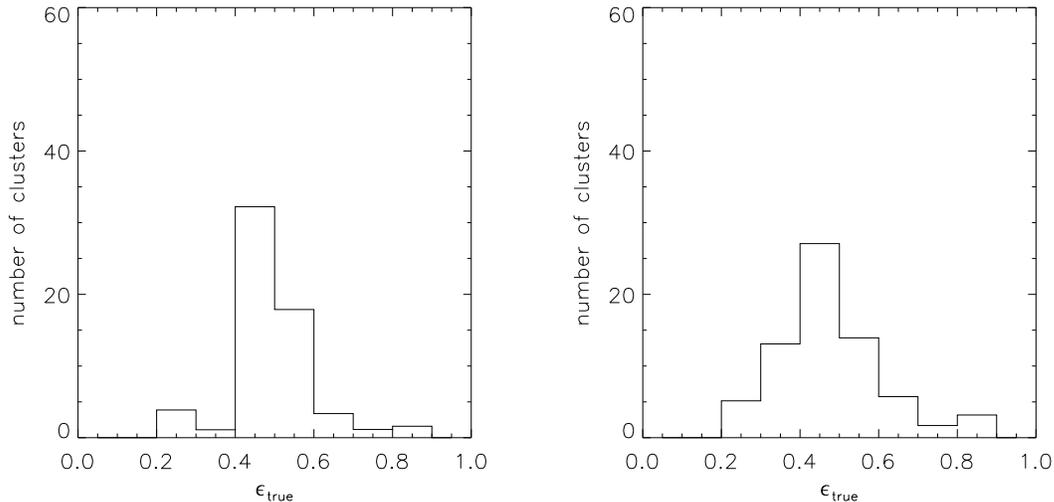

**Figure 4.** Resulting distributions over $\epsilon_{true}$, as calculated by the Lucy iteration, and only showing the signal for $\epsilon_{true,min} < \epsilon_{true} < \epsilon_{true,max}$. A limiting magnitude of 17.0 is adopted. The lefthand figure is for the moments method of RHK, the righthand figure for the tensor method of RHK.

The tensor method of PBF behaves differently from the other two. Especially for small numbers of galaxies per cluster and small $\epsilon_{true}$, the method is noisy. This is clear from Figure 3, where almost no values of $\epsilon_3$ between 0.1 and 0.2 are measured due to discretization effects. For larger $\epsilon_{true}$ and $N_{gal}$ the method performs better, and for high $N_{gal}$ it is similar to the tensor method of RHK. This is to be expected, since both methods basically are the same, except that in the method of PBF the galaxies are binned in cells. For large numbers of galaxies, the discretization due to the binning of the galaxies in cells becomes less significant since more cells are taken into account.

The systematic biases in the three methods can be understood as follows. First, for small $\epsilon_{true}$, the average $\epsilon_{apparent}$ will always be too large since the problem is asymmetric. For example, for $\epsilon_{true} = 0.0$, all values of $\epsilon_{apparent}$ will be larger than or equal to 0.0 and thus the average will also be larger than or equal to 0.0. This effect is analogous to that proposed by Thuan & Gott (1977) to account for the difference between the observed and predicted distributions of axial ratios in elliptical galaxies. For larger $\epsilon_{true}$ this problem becomes less significant. Obviously, for $\epsilon_{true} = 1.0$, a similar problem arises, but now one expects too low an average value. Second, the first two methods use a circular aperture, which tends to make the clusters more spherical. This effect is more important for the more elongated clusters and explains why the measured values are always too low for high $\epsilon_{true}$. For the third method this does not hold, but this method is expected to suffer a lot from discretization noise, since most of the clusters have only 6-8 active cells (PBF; Plionis 1993). This explains the rather discrete behavior of $\epsilon_3$ (Figure 3). Third, the background galaxies also tend to lower the values of $\epsilon_{apparent}$, since they are assumed to be distributed at random over the aperture. The effect will be largest for higher $\epsilon_{true}$ and will have less effect on $\epsilon_2$ than on $\epsilon_1$ and $\epsilon_3$, since in the former method the weight of a galaxy decreases with radius (see Section 3.1).

The PBF method shows some additional features. First, a rather large fraction of the simulated clusters is rejected because they have less than five 'active' cells with five or more galaxies (corresponding to an overdensity of 3.6). Second, almost all rejected clusters have $\epsilon_3 \leq 0.1$. There may even be an effect depending on redshift. At larger redshift, a cluster will in general occupy fewer cells. Therefore, it has a larger chance of being rejected. One may conclude that the spherical clusters are mainly rejected at larger redshift. This effect may be visible in Figure 15(a) of PBF, where the authors find a correlation between redshift and ellipticity, in that the more distant clusters have a larger elongation. They almost find no spherical clusters at larger redshifts.

Note that Fasano et al. (1993) have applied a very similar procedure to determine the shapes of galaxy groups. However, they did not include background galaxies in the simulations that investigate the main errors and biases. Furthermore, the average number of galaxies in groups is only about 10. This will increase the noise in the determination of the shape parameter considerably, as pointed out before.

In the following section the correction cubes will be applied to the distributions of apparent ellipticities (Figure 1).

## 5  CORRECTING THE DISTRIBUTIONS

### 5.1  Lucy's method

What we have is an observed distribution $N(\epsilon_{apparent})$ over projected ellipticities and a probability function $p(\epsilon_{apparent}|\epsilon_{true})$ giving the probability of 'observing' $\epsilon_{apparent}$ if the cluster's true ellipticity is $\epsilon_{true}$. We now want to deduce the distribution of true ellipticities, $N(\epsilon_{true})$, from this. This distribution is given by

$$N(\epsilon_{apparent}) = N(\epsilon_{true})p(\epsilon_{apparent}|\epsilon_{true}) \ . \tag{12}$$

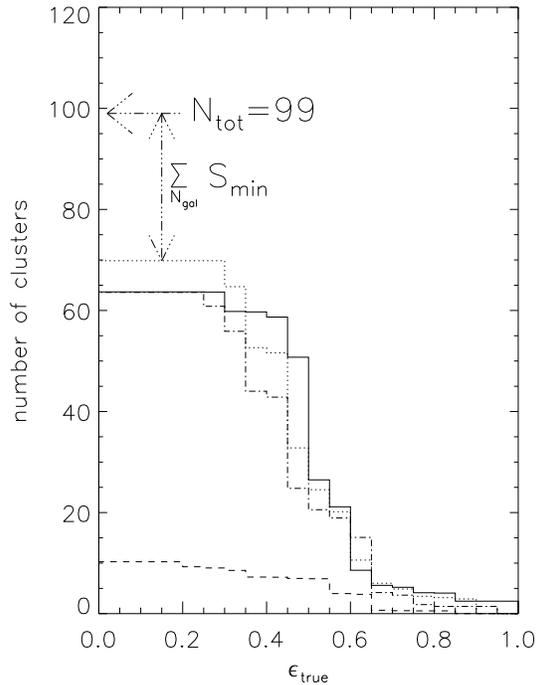

**Figure 5.** Cumulative number of clusters, having a true ellipticity larger than $\epsilon_{true}$. This profile is obtained using the histograms in Figure 4. The solid and the dashed line are for the moments method of RHK at a limiting magnitude of 17.0 and 18.0, respectively. The dotted and dot-dashed line are for the tensor method of RHK at a limiting magnitude of 17.0 and 18.0, respectively. Both the total number of clusters in the sample, $N_{tot}$, and the total number of clusters with $\epsilon_{true} < \epsilon_{true,min}$ are indicated.

However, there is one additional problem: $p(\epsilon_{apparent}|\epsilon_{true})$ depends on the number of galaxies in a cluster. So what one has to do is to bin $N(\epsilon_{apparent})$ according to the number of galaxies per cluster, $N_{gal}$. Then, for each $N_{gal}$-bin, one has a $N(\epsilon_{apparent})$, and one has to solve eq. (12). The standard way to do this is using the Lucy (1974) algorithm. This method guarantees smooth, positive solutions. However, the inversion from $N(\epsilon_{apparent})$ to $N(\epsilon_{true})$ will be non-unique, as is obvious from Figure 3, because $\partial \epsilon_{apparent}/\partial \epsilon_{true} \sim 0$ for $\epsilon_{true}$ smaller than some critical value $\epsilon_{true,min}$. For $\epsilon_{true} < \epsilon_{true,min}$, all $N(\epsilon_{true})$ give the same contribution to $N(\epsilon_{apparent})$, so one can not discriminate between different $N(\epsilon_{true})$-profiles in this $\epsilon_{true}$-range. The same effect operates near $\epsilon_{true} \sim 1.0$ because there also $\partial \epsilon_{apparent}/\partial \epsilon_{true}$ can become zero. If this happens, there is a critical $\epsilon_{true,max}$, which is in general close to 1.0. Also, only few clusters are expected to have such large $\epsilon_{true}$. Note that the values of $\epsilon_{true,min}$ and $\epsilon_{true,max}$ depend on the magnitude limit one adopts, the method used, and the $N_{gal}$-bin.

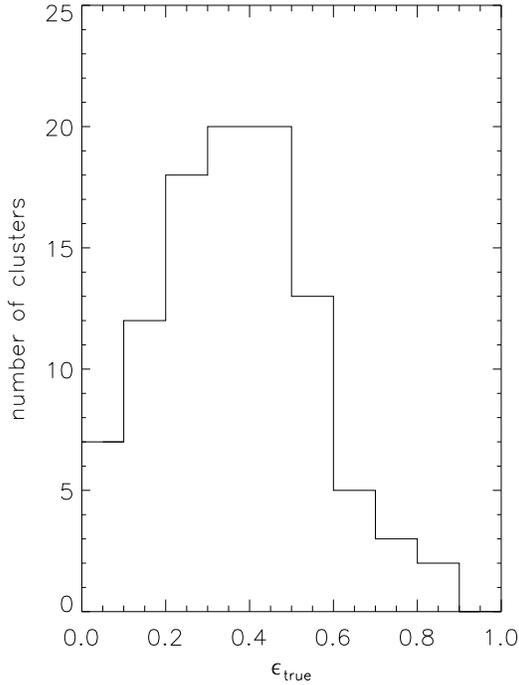

**Figure 6.** Distribution over true projected ellipticities.

**Table 1.** Number of clusters that have either $\epsilon_{true} < \epsilon_{true,min}$ or $\epsilon_{true} > \epsilon_{true,max}$. A limiting magnitude of 17.0 is adopted. The total number of clusters in the sample is 99, though method 3 rejects more than one-third or these because they occupy too few active cells.

| method 1 | | method 2 | | method 3 | |
|---|---|---|---|---|---|
| $\epsilon_{true,min}$ | $S_{min}$ | $\epsilon_{true,min}$ | $S_{min}$ | $\epsilon_{true,min}$ | $S_{min}$ |
| 0.4 | 2.64 | 0.4 | 10.60 | 1.0 | 57.86 |
| 0.3 | 32.26 | 0.3 | 18.47 | 0.2 | 0.99 |
| 0.2 | 0.11 | | | | |
| | | | | | |
| $\epsilon_{true,max}$ | $S_{max}$ | $\epsilon_{true,max}$ | $S_{max}$ | $\epsilon_{true,max}$ | $S_{max}$ |
| 0.9 | 2.47 | | | | |

For each $N_{gal}$-bin we do the Lucy-iteration, and after convergence one is left with a distribution over $\epsilon_{true}$ in the range $\epsilon_{true,min} \leq \epsilon_{true} \leq \epsilon_{true,max}$, with the total number of clusters having $\epsilon_{true} < \epsilon_{true,min}$ given by

$$S_{min} = \int_{0.0}^{\epsilon_{true,min}} N(\epsilon_{true}) d\epsilon_{true} \quad , \tag{13}$$

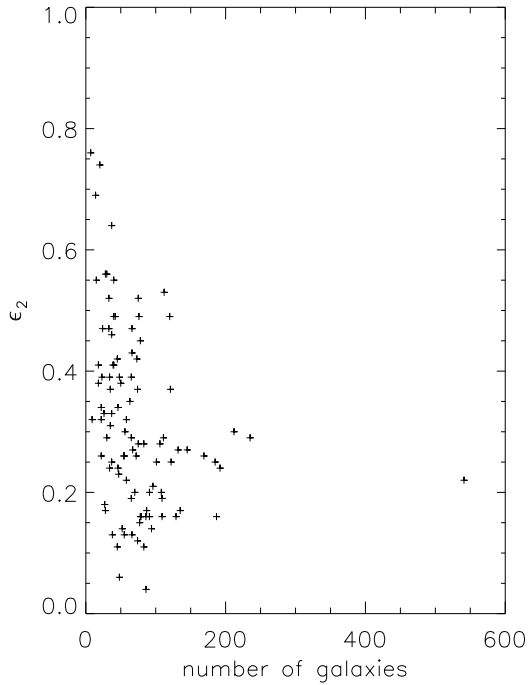

**Figure 7.** Ellipticity versus the number of galaxies in a cluster. The ellipticity is the apparent ellipticity as measured by the tensor method of RHK and as applied to the original data. A limiting magnitude of 17.0 was adopted.

and with the total number of clusters having $\epsilon_{true} > \epsilon_{true,max}$ given by

$$S_{max} = \int_{\epsilon_{true,max}}^{1.0} N(\epsilon_{true}) d\epsilon_{true} \quad . \tag{14}$$

Table 1 gives $S_{min}$ and $S_{max}$ for all three methods and adopting a limiting magnitude of 17.0. $S_{min}$ and $S_{max}$ generally increase when adopting a larger limiting magnitude. This is because the number of background galaxies roughly increases by $10^{0.6 m_{lim}}$ (Binggeli 1982), i.e., a factor of $\sim 4$ per magnitude, while the number of cluster galaxies increases less rapidly. Then, we combine all $N_{gal}$-bins by simply summing all $N(\epsilon_{true})$ in the intervals $\epsilon_{true,min} \leq \epsilon_{true} \leq \epsilon_{true,max}$. The results of this summation are shown in Figure 4. The lefthand panel is for the moments method of RHK, the righthand panel for the tensor method of RHK. These diagrams thus show the contribution to the distribution over true projected ellipticities, resulting from the Lucy-iteration in the intervals where $\epsilon_{true}$ is uniquely determined. For the PBF-method, no diagram is shown, because this would be completely empty. That is, at a limiting magnitude of 17.0, the number of galaxies in the clusters is so small that the PBF-method has no intervals in $\epsilon_{true}$ where it can be uniquely determined.

Now one has to take into account the number of clusters $S_{min}$ and $S_{max}$. For each $N_{gal}$-bin, we have the total number of clusters that have $\epsilon_{true} < \epsilon_{true,min}$, or that have $\epsilon_{true} > \epsilon_{true,max}$.

What one can do is plot the number of clusters that have a true ellipticity larger than a certain $\epsilon_{true}$. This cumulative distribution is shown in Figure 5, for both methods of RHK, and at limiting magnitudes of 17.0 and 18.0. The solid and dashed lines show the results for the moments method at limiting magnitudes 17.0 and 18.0, respectively. The dotted and dot-dashed lines show the results for the tensor method at limiting magnitudes 17.0 and 18.0, respectively. As $S_{min}$ is only non-zero below a certain $\epsilon_{true,min}$ (which depends on limiting magnitude and method), all profiles should coincide above this value. For example, at $m_{lim} = 17.0$, both methods only have a non-zero contribution from $S_{min}$ below $\epsilon_{true} = 0.4$ (the number $S_{max} = 2.47$ for the first method has already been accounted for in Figure 5). Therefore, for $\epsilon_{true} > 0.4$ the solid and dotted line should coincide. This is not exactly the case, but it is very well possible to select a best-fitting profile to all curves. Note that the dashed line (moments method, $m_{lim} = 18.0$) shows that the number of clusters in all ($\epsilon_{true,min}, \epsilon_{true,max}$)-intervals drastically decreases when the limiting magnitude is increased (as stated before). This curve hardly puts any constraints on the total distribution of clusters over $\epsilon_{true}$. Then, for a specific method and a certain $m_{lim}$, all contributions of $S_{min}$ have to be added to the profiles in Figure 5, under the constraints that the curves should be consistent and that the total number of clusters is 99 (as indicated in the figure). This 'extrapolation' is performed with a very low order polynomial fit. That is, we assume there to be no rapid variations in the number of clusters at low $\epsilon_{true}$. That it is indeed possible to make all curves nearly consistent guarantees that one gets a nearly unique solution.

### 5.2   Results

The resulting distribution over $\epsilon_{true}$ is shown in Figure 6, and is just the derivative of the 'extrapolated' profiles in Figure 5. It looks rather different from the apparent distributions of RHK in Figure 1. Not only does the peak shift to $\epsilon \sim 0.4$, but the profile also extends to $\epsilon \sim 0.8$, whereas in Figure 1 there were no clusters with elongations larger than $\epsilon \sim 0.4$. This suggests that clusters are more elongated than the RHK-methods in Figure 1 would indicate. The PBF method underestimates the ellipticities much less, but this is partly due to discretization.

To check the results, we construct 99 clusters satisfying both the profile of Figure 6 and the distribution over number of galaxies per cluster as in the original RHK data set. Then we apply all three methods to these data and compare them to the original observations. To include a possible relation between the ellipticity and number of galaxies of a cluster, the apparent ellipticities (determined with the tensor method of RHK and for $m_{lim} = 17.0$) are plotted versus the number of galaxies in the cluster (Figure 7). Although the uncorrected ellipticities in Figure 7 will not be correct quantitatively, the following can be concluded: because for clusters containing many galaxies, $\epsilon$ is rather well determined, the fact that $\epsilon_2 < 0.3$ for $N_{gal} > 150$ indicates that there are no clusters in this $N_{gal}$-range which have $\epsilon_{true}$ larger than 0.45 (as obtained from the correction cube). The clusters with fewer than ~150 galaxies show measured ellipticities up to ~0.8. Because the true ellipticity is almost always larger than the apparent one, and because there really are clusters with $\epsilon_{true} > 0.45$ (see Figure 6), this means that these clusters (with $N_{gal} < 150$) must have rather large true ellipticities. Thus, there is a clear anti-correlation between cluster ellipticity and its number of galaxies (see also Section 5.3). The same anti-correlation is also found using the apparent ellipticities of the other two methods and limiting magnitudes of 18.0 and 19.0. For the method of PBF, the anti-correlation weakens towards lower limiting magnitudes, and for a limiting magnitude of 17.0 it is almost completely absent, as noted by PBF in their Figure 15b. But this is due to noise and discretization, as discussed in Section 4.3.

Table 2. Assumed anti-correlation between cluster elongation and its number of galaxies for the different limiting magnitudes.

| $\epsilon_{true}$ | $N_{gal}$ ($m_{lim} = 17.0$) | $N_{gal}$ ($m_{lim} = 18.0$) | $N_{gal}$ ($m_{lim} = 19.0$) |
|---|---|---|---|
| [0.0, 0.2] | [150, 500] | [200, 1000] | [200, 1000] |
| [0.2, 0.4] | [50, 200] | [100, 300] | [100, 350] |
| [0.4, 1.0] | [0, 100] | [0, 150] | [0, 200] |

The anti-correlation we used in assigning an ellipticity and number of galaxies to the Monte-Carlo clusters is listed in Table 2. The Monte-Carlo clusters are constructed according to the result in Figure 6 and the distribution over number of galaxies per cluster from the original RHK data. The apparent elongations can be compared with the original data.

The results are shown in Figures 8 and 9. In Figure 8 the lefthand column shows the profiles as obtained directly from the data set of RHK. The righthand column shows the $\epsilon_{apparent}$ distributions as obtained from the simulations. The limiting magnitudes are 17.0, 18.0, and 19.0 (from top to bottom). Again, the solid line is for the moments method of RHK, the dotted line for the tensor method of RHK, and the dashed line for the tensor method of PBF. Note that the lower lefthand plot is somewhat different from Figure 1 because the centers of the clusters were redetermined (see Section 2.1). The agreement between real observations and the predicted 'observations' is good for both RHK methods and at all three magnitude limits. For the PBF method the agreement is somewhat worse. The Kolmogorov-Smirnov probability for the observations and predictions to be the same is always larger than 0.97, except for $\epsilon_3$ at a limiting magnitude of 17.0 (KS probability is 0.83) and at a limiting magnitude of 19.0 (KS probability is 0.00). This, again, illustrates that the PBF method is unable to reproduce the original results because of the rather large scatter in the method. Figure 9 can be compared with Figure 7. The agreement is very good. Although the assumed anti-correlation described in Table 2 is rather simple, it is necessary to reproduce both the distributions over $\epsilon_{apparent}$ and the correlation between $\epsilon_{apparent}$ and $N_{gal}$ in Figure 7. This indicates that the obtained profile over true projected ellipticities in Figure 6, together with the assumed anti-correlation between ellipticity and number of galaxies of Table 2, are at least consistent with the observations. Furthermore, it is clear that a $\epsilon_{true}$-distribution, which does not depend on magnitude, as is assumed in constructing the distribution over $\epsilon_{true}$ (see Section 5.1), can also explain the differences in observed profiles for different $m_{lim}$, i.e., while the observations show that the average ellipticity of a cluster increases for the more luminous galaxies, this may only be an artifact of the methods and due to the smaller number of galaxies included. The upper row of Figure 8 also show a property of the PBF-method, mentioned previously. For lower limiting magnitudes this method rejects almost half of the clusters because they do not occupy 5 active cells. Also, it shows the discretization noise of this method: for high limiting magnitudes the result is similar to that of the tensor method of RHK, but for lower magnitudes it gets inconsistent and shows an almost flat distribution.

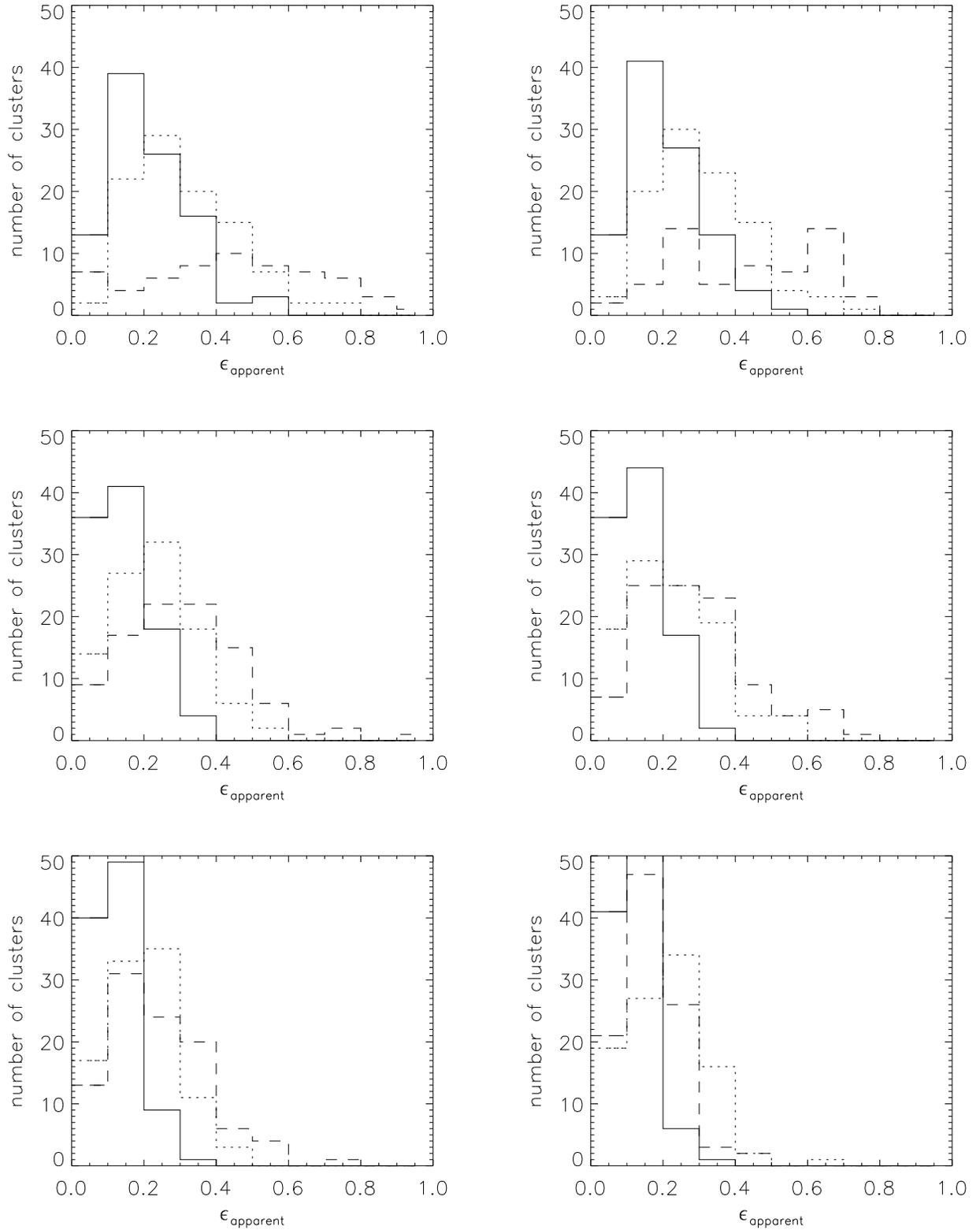

**Figure 8.** Results of simulating the data. The lefthand column shows the observed profiles over ellipticity, as calculated directly from the RHK data set (using the uncorrected ellipticities). The righthand column shows the 'observed' profiles over ellipticity, as reconstructed from the true profile of Figure 5 and from the distribution over numbers of galaxies. From top to bottom, the plots show the results for limiting magnitudes of 17.0, 18.0, and 19.0, respectively. The solid line is for the moments method of RHK, the dotted line is for the tensor method of RHK, and the dashed line is for the tensor method of PBF.

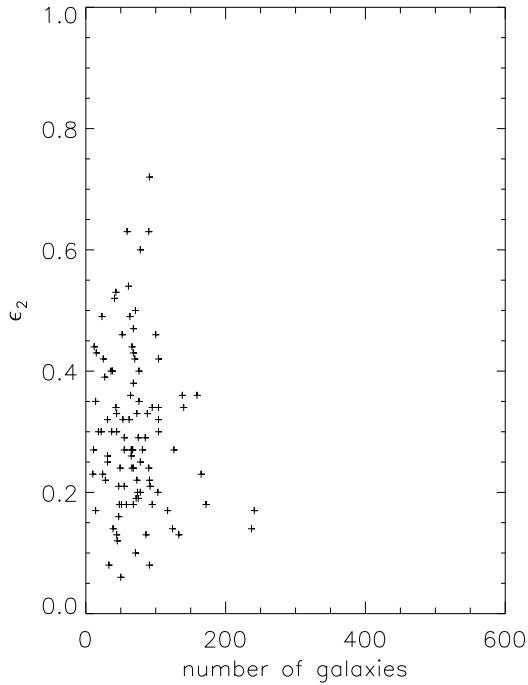

**Figure 9.** Ellipticity versus the number of galaxies in a cluster. The ellipticity is the apparent ellipticity as measured by the tensor method of RHK and applied to the model clusters. A limiting magnitude of 17.0 was adopted.

The errors in Figure 8 are rather small and are obtained by trying different distributions over $\epsilon_{true}$ and different anti-correlation schemes which all produce the same profile over $\epsilon_{apparent}$ as originally observed, and which show the same relation between ellipticity and number of galaxies (see Figure 7). The errors are largest in the lowest bins. This is to be expected, since at low $\epsilon_{true}$, the measured quantity will be always very similar (see Figure 3).

Our results can be compared with earlier work of e.g. Carter & Metcalfe (1980) and Binggeli (1982). Carter & Metcalfe showed clusters to have an average ellipticity of $0.44 \pm 0.04$, somewhat larger than ours. However, their sample consisted of only 21 clusters and they calculated $\epsilon$ at $0.5h^{-1}$ Mpc. It is obvious from Figure 2 that $\epsilon_{apparent}$ depends on the radius at which it is evaluated. The authors indeed found a decrease in ellipticity outwards. This was also noted by Flin (1984) and Burgett (1982). More inwards, there is a larger scatter due to the smaller number of galaxies involved. Binggeli (1982) showed the clusters to have high elongations with a peak at $\epsilon \sim 0.45$ and extending to 0.8. However, he only used 44 clusters each with only 50 galaxies, thereby increasing the noise considerably. Thus, although the previous studies are consistent with our results, this study has much better statistics and shows internal consistency: we are able to reproduce all features in the observations: an apparent increase in ellipticity for the more luminous galaxies, and an anti-correlation between ellipticity and number of galaxies in the cluster.

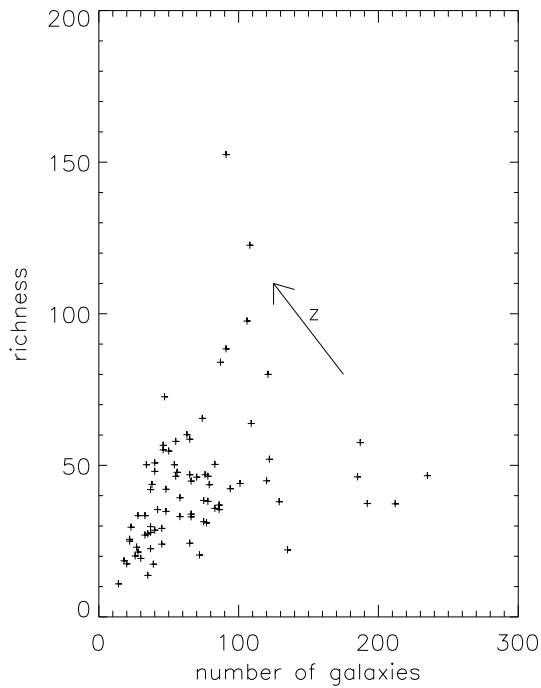

**Figure 10.** Relation between the richness of a cluster, as defined by Vink & Katgert (1994), and its number of galaxies within $1h^{-1}$ Mpc (evaluated at a limiting magnitude of 17.0).

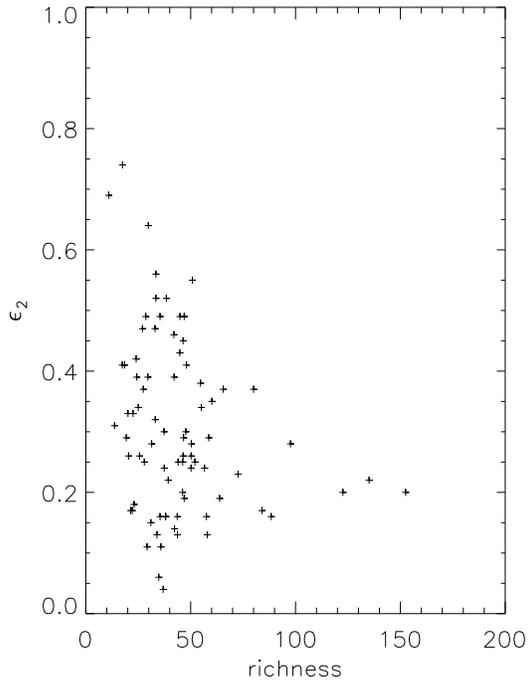

**Figure 11.** Correlation between the apparent ellipticity of a cluster and its (redshift-independent) richness. The former is measured by the tensor method of RHK and for a limiting magnitude of 17.0.

### 5.3 Relation between $\epsilon$ and richness

We have argued before (see Section 5.2) that the clusters containing more galaxies tend to be more spherical. However, an obvious bias is observed, in that the clusters containing the most galaxies, are most nearby. So it is not clear whether these clusters are intrinsically rich, or whether they just appear rich. To investigate this, we plot the richness of a cluster versus its number of galaxies within $1h^{-1}$ Mpc and for a limiting magnitude of 17.0 (Figure 10). Here the richness is defined by Vink & Katgert (1994) as the number of cluster galaxies within the absolute magnitude interval [-22.5,-21.5]. This number is corrected for background contamination. This definition of richness makes it independent of redshift, in contrast to the original definition of Abell (1958). From Figure 10 it is obvious that there is a clear correlation between the richness of a cluster and its number of galaxies. Clusters of fixed redshift lie approximately on a straight line through the origin. For increasing redshift this line steepens in the direction of the arrow. For fixed richness, clusters at lower redshift have a higher number of galaxies.

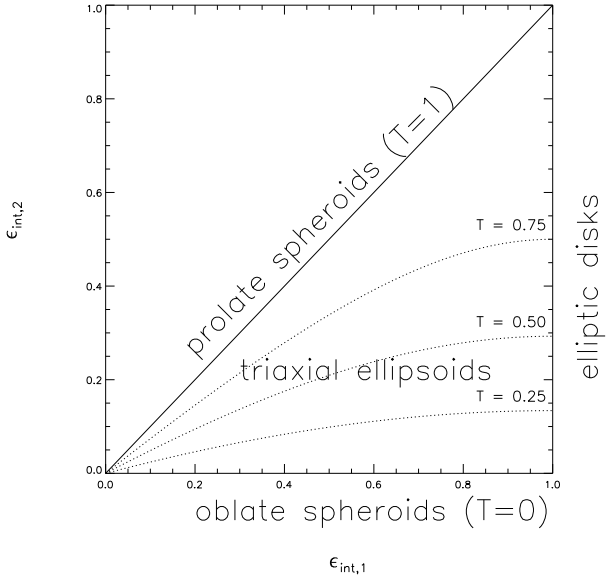

**Figure 12.** Plane of intrinsic ellipticities ($\epsilon_{int,1}, \epsilon_{int,2}$). Only shapes with $\epsilon_{int,2} \leq \epsilon_{int,1}$ are possible. Dotted lines are lines of constant triaxiality $T$. Oblate and prolate spheroids correspond to $T=0$ and $T=1$, respectively.

Together with the result of Figures 7 and 9 this means that rich clusters are intrinsically more spherical than poor ones (as we have assumed in simulating the observations in Section 5.2). This is shown in Figure 11 where we plot the apparent ellipticity of a cluster versus its (redshift independent) richness. The correlation is even more clear than in Figure 7. Even though the values of $\epsilon_2$ will not be correct quantitatively, as in Figure 7, qualitatively the trend of this figure is correct. The same correlation was found by Struble & Ftaclas (1994) who calculated the shapes of 350 clusters of galaxies as a function of their shape and velocity dispersion.

The physical reason for this correlation may be that regions of higher density turn around earlier from the Hubble flow than lower density regions. Since we have a fixed aperture of $1h^{-1}$ Mpc, richer clusters will have a higher overdensity inside this aperture. The mean relaxation time will also be shorter and this will cause the more spherical shape for rich clusters. This result is in disagreement with the result of PBF, who found no correlation between the shape of a cluster and its number of galaxies (their Figure 15a).

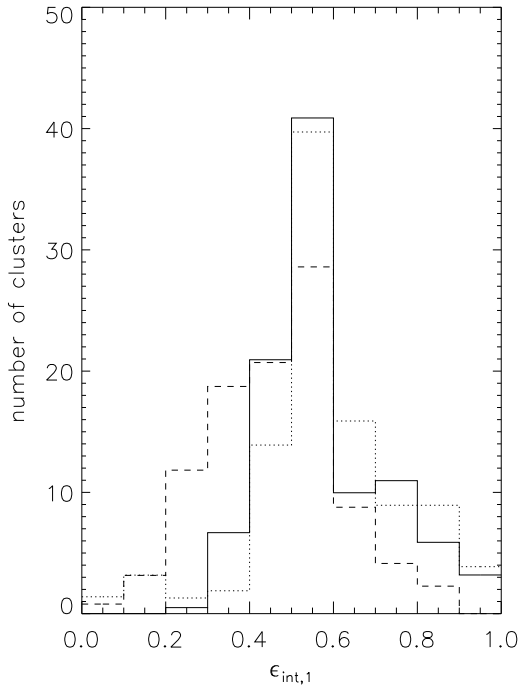

**Figure 13.** Inverted distribution of intrinsic ellipticities for three different values of the triaxiality parameter $T$. The solid line is for $T = 0.0$ (oblate), the dotted line is for $T = 0.5$ (intermediate triaxial), and the dashed line is for $T = 1.0$ (prolate).

## 6 INTRINSIC SHAPES

Since the projected ellipticity results from the projection of two intrinsic ellipticities, the problem of determining the distribution over the two intrinsic ellipticities is underconstrained. One possible way out is to assume all clusters to have the same triaxiality parameter, $T$, defined as

$$T = \frac{a^2 - b^2}{a^2 - c^2} = \frac{\epsilon_{int,2}(2 - \epsilon_{int,2})}{\epsilon_{int,1}(2 - \epsilon_{int,1})} \quad , \tag{15}$$

where $a, b, c$ are the axes of the clusters (see eq. [8]) and $\epsilon_{int,1} = 1-(c/a)$ and $\epsilon_{int,2} = 1-(b/a)$ the intrinsic ellipticities, with $\epsilon_{int,1}$ the larger of the two. This procedure was also used by Franx, Illingworth & de Zeeuw (1991), who applied it to elliptical galaxies.

Figure 12 shows the plane of all possible shapes ($\epsilon_{int,1}$, $\epsilon_{int,2}$). Curves of constant $T$ are shown. The limiting cases $T = 0$ and $T = 1$ correspond to oblate and prolate shapes, respec-tively. We calculate the distribution over intrinsic ellipticities with Lucy's (1974) method, assuming all clusters to have the same triaxiality parameter $T$. This method guarantees smooth, positive solutions. The resulting distributions over $\epsilon_{int,1}$ are shown in Figure 13 for $T = 0.0$ (oblate models, solid line), 0.5 (intermediate triaxial models, dotted line), and 1.0 (prolate models, dashed line). All the distributions peak at $\epsilon_{int,1} \sim 0.55$. In the oblate case, no clusters with $\epsilon_{int,1} \sim 0.0$ occur, and a rather large percentage has a very large elongation. As expected, the purely prolate distribution extends to smaller $\epsilon_{int,1}$ and shows less very elongated clusters. This is because a prolate cluster has a rather small probability to be viewed as being spherical, so many of them are needed to produce the signal at small projected ellipticities. Oblate clusters have a rather small chance of being observed very elongated, so many of these are needed to produce the signal at very large projected ellipticities. The $T = 0.5$ case is intermediate.

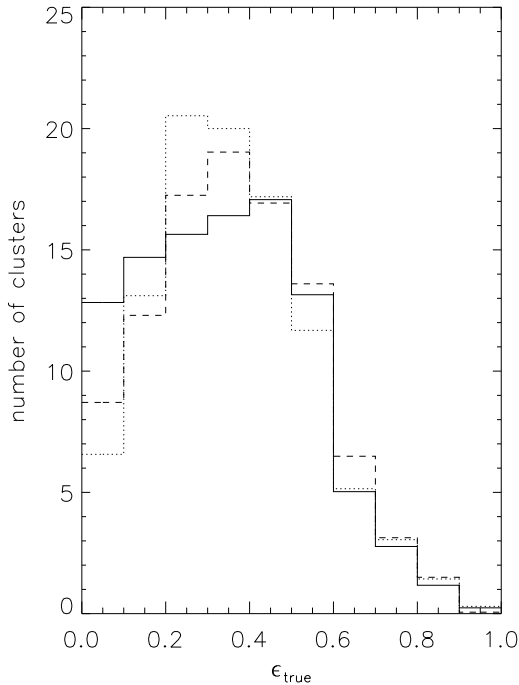

**Figure 14.** Recovered distribution of true ellipticities for three different values of the triaxiality parameter $T$. The solid line is for $T = 0.0$ (oblate), the dotted line is for $T = 0.5$ (intermediate triaxial), and the dashed line is for $T = 1.0$ (prolate).

The Lucy iteration always gives a best fitting solution, and one does not yet know how good this fit is. To check this, we generate Monte-Carlo clusters according to the profiles in Figure 13 and then reconstruct the distribution over $\epsilon_{true}$ from this. The results are shown in Figure 14 and should be compared with Figure 6. For large elongations, all $T$-values are equally good. However, the oblate distribution produces too many (nearly) spherical clusters. The other two distributions perform equally well. It is therefore to be expected that most of the clusters will have a triaxiality parameter between $\sim 0.5$ to $1.0$.

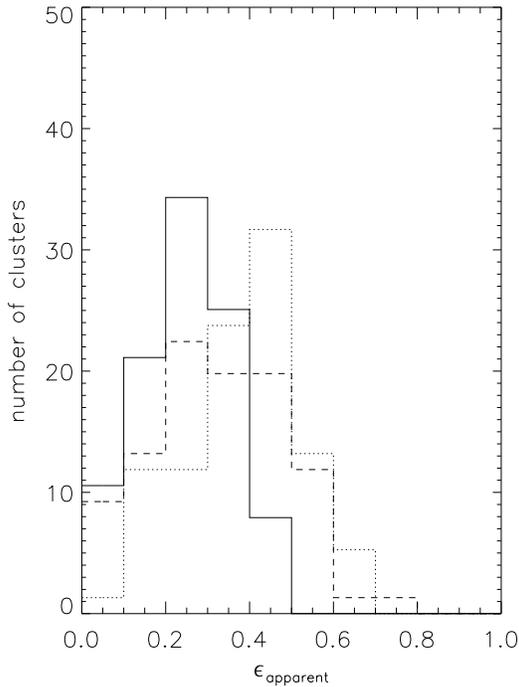

**Figure 15.** Distribution over observed ellipticities for clusters simulated in an $\Omega = 1.0$ universe. The lines have the same meaning as in Figures 1 and 7.

The following has to be remarked with regard to the intrinsic distributions. Both for a purely prolate and a purely oblate distribution there is a bias in the projected distribution because Abell clusters are defined as an overdensity in the surface density of galaxies, For prolate clusters, a cluster with its long axis pointing towards us has a larger probability of being detected since its surface density will in general be higher than that of a cluster with its short axis pointing along the line of sight. This means that for prolate clusters there is a bias towards observing more spherical than elongated ones. In this case, the projected distribution has to be modified by decreasing the number of low-$\epsilon_{true}$ clusters. This will also change the intrinsic distribution slightly. In the case of a purely oblate distribution, the opposite holds. An oblate cluster has a larger probability of being detected if its short axis is in the plane of the sky. This produces a bias towards more elongated clusters and one has to modify the distribution over projected ellipticities by increasing the number of low-$\epsilon_{true}$ clusters. For the intermediate triaxial clusters ($T = 0.5$), the bias is much less, and the $\epsilon_{true}$ distribution has to modified only slightly.

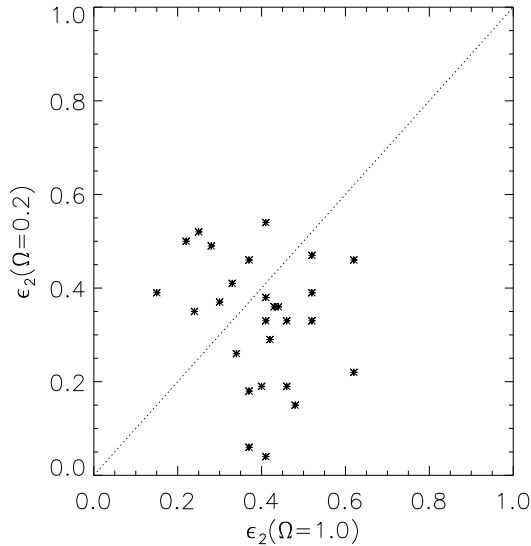

**Figure 16.** Relation between the ellipticity of a cluster generated in an $\Omega = 1.0$ universe and a cluster generated in an $\Omega = 0.2$ universe, using the same random number seed to generate the clusters. For all nine clusters the ellipticities are calculated for the xy-, xz-, and yz-projection, and using the tensor method of RHK.

## 7 N-BODY SIMULATIONS

As mentioned in the Introduction, the shapes of clusters will most certainly be influenced by the density parameter $\Omega$. For low $\Omega$, structures will collapse at a higher redshift than in an $\Omega = 1.0$ universe (Maoz 1990) and thus have had more time to virialize and wipe out their asymmetries and substructures. Furthermore, the timescale in a low-$\Omega$ universe is longer than in an Einstein-de Sitter universe (see e.g. Padmanabhan 1993, eq. [2.79]). What may be also important is that in low-$\Omega$ case, there is only little secondary infall at late times. This means that perturbations in a low $\Omega$ universe cease to grow. On the contrary, $\Omega = 1.0$ clusters keep accreting matter and this may change their structure considerably.

In order to try and put constraints on $\Omega$, we have used N-body simulations of clusters in a CDM universe (van Kampen 1994). These simulations use a volume of radius $25-32h^{-1}$ Mpc, which is large enough to contain the turn-around radius and to model tidal fields. At a certain epoch galaxies are identified using a local density percolation technique combined with a virial equilibrium condition. For each galaxy found its constituent particles are replaced by a single softened particle, where the binding energy of the original particles is transferred into internal energy of the single particle galaxy. That the galaxy identification algorithm can be important was shown by Summers (1993). The projected density of the clusters on the 'sky' satisfies that of a modified Hubble-profile (eq. [10]), and also the number of galaxies is comparable to the original data. For a complete description of the models and method see van Kampen (1994).

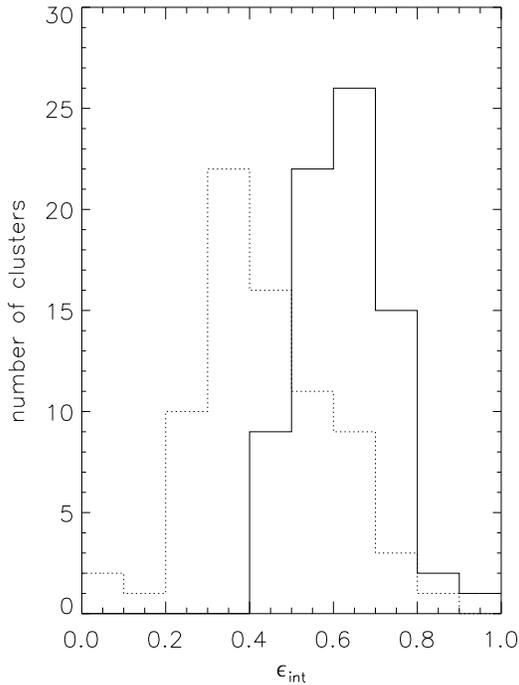

**Figure 17.** Distribution over intrinsic ellipticities $\epsilon_{int,1}$ (solid line) and $\epsilon_{int,2}$ (dotted line) for the simulated clusters with $\Omega = 1.0$. The ellipticities are calculated with the tensor method of RHK.

**Table 3.** Mean ellipticities of the three orthogonal projections of the nine clusters, generated both using $\Omega = 1.0$ and $\Omega = 0.2$, and calculated with the three methods. The numbers in parentheses give the dispersions in $\epsilon_i$.

|  | $\Omega = 1.0$ | $\Omega = 0.2$ |
| --- | --- | --- |
| $\langle \epsilon_1 \rangle$ | 0.27 (0.01) | 0.22 (0.01) |
| $\langle \epsilon_2 \rangle$ | 0.40 (0.01) | 0.33 (0.02) |
| $\langle \epsilon_3 \rangle$ | 0.26 (0.04) | 0.29 (0.02) |

For $\Omega = 1.0$, 75 clusters are generated according to the probability distribution of the parameters $\nu$, $x$, $e$ and $p$, as given by Bardeen et al. (1986). These parameters determine the distribution of heights ($\nu$), curvature ($x$) and shapes ($e$ and $p$) of the peaks in the initial density field, out of which structures are assumed to form. A biasing parameter $b = 2.0$ was adopted for the $\Omega = 1.0$ catalogue. Using a 2-D identification procedure, 4 out of the

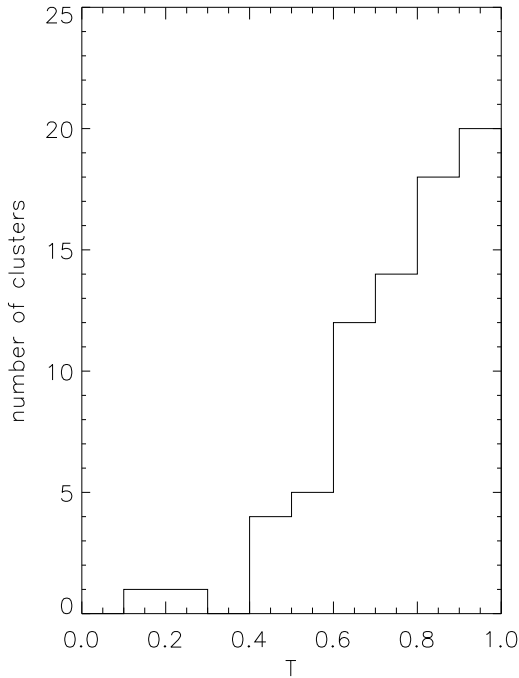

**Figure 18.** Distribution over the triaxiality parameter T for the N-body simulated clusters with $\Omega = 1.0$. The ellipticities are taken from Figure 16.

75 clusters are found to produce Abell-like clusters of richness class 0. All others model clusters resemble Abell clusters of richness $R \geq 1$. For this choice of the bias parameter the catalogue of simulated clusters is thus expected to be a fair representation of real clusters in the universe. Then each cluster is viewed from a random direction. All galaxies in the simulation cylinder that are projected onto the cluster area are used in order to include some background and foreground. For the two methods of RHK, we only use the galaxies inside a projected distance of $1h^{-1}$ Mpc of the cluster center. This center is determined in the same way as was done for the real data.

The distribution over uncorrected (=apparent) ellipticities is plotted in Figure 15. As before, the solid lines indicate the moments method of RHK, the dotted lines denote the tensor method of RHK, and the dashed lines are the results for the tensor method of PBF. This figure can be compared with Figure 1. It is obvious that the ellipticities of the simulated clusters are too large to be consistent with the observations. To examine the influence of $\Omega$ on the shapes of the clusters, we have selected nine clusters (both massive and less massive) from the $\Omega = 1.0$ catalogue. These, then, are again simulated in an unbiased $\Omega = 0.2$ universe using the same random number seeds. The nine $\Omega = 0.2$ clusters are a representative sample of a catalogue being constructed at present, analogous to the catalogue of $\Omega = 1.0$ clusters. The properties of this whole catalogue will be presented elsewhere. For each of these eighteen clusters (i.e. nine clusters both in $\Omega = 1.0$ and $\Omega = 0.2$) we have calculated $\epsilon_2$ for the three orthogonal projections and using the tensor method of RHK (method 2). The mean values of $\epsilon_{apparent}$ for all three methods are shown in Table 3. The numbers in parentheses give the dispersions. The relations between $\epsilon_2$ in the $\Omega = 1.0$ and the $\Omega = 0.2$ simulations are shown in Figure 16. From this figure it is clear that clusters in an $\Omega = 0.2$ universe tend to be more spherical than those in an $\Omega = 1.0$ universe. Examining all clusters individually, it turns out that almost all of the data having $\epsilon_{apparent}(\Omega = 0.2) > \epsilon_{apparent}(\Omega = 1.0)$ correspond to clusters that either have just merged in the $\Omega = 1.0$ simulation and therefore appear more spherical in this simulation, or that are double clusters or appear to be due to the projection of two groups onto each other. All 'isolated' clusters turn out to lie below the dotted line. When the whole $\Omega = 0.2$ catalogue will be finished, we will have a handle on the fraction of clusters that are merging at present or that are/appear double. With this we will be able to put more quantitative constraints on the effect visible in Figure 16 and therefore on $\Omega$.

Bardeen et al. (1986) show that the shapes of the density perturbations in the initial density field, which we expect to correspond with what we now call clusters, are somewhat more elongated for lower $\Omega$ models. That their shapes are now less elongated points at a quite different evolution of clusters for different $\Omega$. For low $\Omega$, clusters collapse at a larger redshift, time elapses faster with redshift, and clusters do not accrete much material from outside during the last epochs. The latter means that $\Omega = 0.2$ clusters nearly have not grown during the last epochs and therefore their virialization process is not disturbed by clumps of infalling matter.

It should be noted that these simulations show the same features as the original RHK data in that a cluster looks more elongated when considering only the brighter (i.e. more massive) galaxies. As mentioned previously, this is just an effect of the methods, and is not likely to be real. Also, the PBF method rejects a fraction of the clusters. Furthermore, clusters containing more galaxies are also intrinsically more spherical, though the effect is less pronounced than in the real data.

To see how reliable it is to assume all clusters to have the same triaxiality parameter $T$ (as assumed in Section 6), we determine the two (apparent) intrinsic ellipticities $\epsilon_{int,1}$ and $\epsilon_{int,2}$ for the complete $\Omega = 1.0$ catalogue. The ellipticities are calculated using the three-dimensional analog of the tensor method of RHK and the calculations are done within a (three-dimensional) radius of $1h^{-1}$ Mpc, to resemble as good as possible the projected distribution. The results are shown in Figure 17. The solid line denotes the distribution over $\epsilon_{int,1}$, and the dotted line indicates $\epsilon_{int,2}$. The distribution over $T$ is shown in Figure 18. It is clear that more than 50 % of the clusters have a triaxiality parameter T larger than 0.8, which means that they are nearly prolate. Almost no clusters have $T < 0.4$, which means that there are almost no oblate clusters in this sample. This result, although determined for the $\Omega = 1.0$ clusters and using the apparent ellipticities, is fully consistent with our analysis of the real clusters, where we found that the distribution over true ellipticities is better matched by a purely prolate distribution than a purely oblate one. It is not expected that this distribution will change drastically if we use the true intrinsic ellipticities instead of the apparent ones, because both $\epsilon_{int,1}$ and $\epsilon_{int,2}$ will then change, and the effect on $T$ may be small.

## 8  DISCUSSION AND CONCLUSIONS

We have reananlyzed a data set of 99 Abell clusters data, studied previously by Rhee, van Haarlem and Katgert (1989). The clusters are of Abell richness $R \geq 1$, redshift $z \leq 0.1$, $10^h \leq \alpha \leq 18^h, \delta < -25^o$ and $b \geq 30^o$. This data set is complete up to $z = 0.08$. To determine the shapes of the clusters, the same methods that these authors applied are used. Also, a method that has been used by Plionis, Barrow & Frenk (1991) in their analysis of the Lick catalogue, is applied. However, these three methods are found to have a number of errors and biases associated with them. By simulating clusters with a Monte-Carlo technique (using a Hubble-profile with a core radius of $0.25h^{-1}$ Mpc), these errors and biases are investigated. In fact, the adopted core radius is found not have a large influence in the range $0.1 - 0.5h^{-1}$ Mpc. Also the exact number of background galaxies adopted is of little influence.

It is found that for small (projected) ellipticity $\epsilon_{true}$ (i.e., large axis ratio), the methods overestimate the ellipticity. This is due to the asymmetrical properties of the problem and is analogue to the Gott & Thuan effect (Thuan & Gott 1977). For large $\epsilon_{true}$, the methods underestimate its value due to the background and the spherical aperture (in the case of the Rhee et al. methods).

The observed distributions over cluster shapes are corrected for the errors and biases, using the iterative Lucy-algorithm. The corrected distribution shows a peak at $\epsilon_{true} \sim 0.4$ and extends to $\sim 0.8$. The number of nearly spherical clusters (with $\epsilon_{true} \leq 0.2$) is limited to about 20%. These results are consistent with some of the previous studies, but have much better statistics due to both the larger number of clusters in our sample and the larger number of galaxies per cluster. Furthermore, the result is internally consistent. That is, we are able to reproduce both the observed distributions over $\epsilon_{apparent}$ and the observed relation between $\epsilon_{apparent}$ and the number of galaxies in a cluster. This is only possible if one assumes an anti-correlation between the number of galaxies in a cluster and its *true* ellipticity. Using redshift-independent richnesses of Vink & Katgert (1994), defined as the number of galaxies in the absolute magnitude interval [-22.5,-21.5], we show that the number of galaxies in a clusters is correlated with the richness of that cluster and is not just an effect of the cluster distance. This means that richer clusters are intrinsically more spherical than poor ones. It may point to the fact that rich clusters have turned around from the Hubble-flow earlier and have had more time to virialize. Also, the virialization process will be faster due to the larger number of galaxies. To achieve the internal consistency, it is not necessary to assume that clusters are more elongated if only the brighter galaxies are included (as was suggested by Binney 1977). This turns out to be only an artifact due to the smaller number of galaxies used to determine $\epsilon_{apparent}$.

We try to say something about the intrinsic shapes of clusters by assuming that they all have the same triaxiality parameter $T$, as defined by Franx, Illingworth & de Zeeuw (1991). If one assumes all clusters to be oblate (i.e., have $T = 0.0$), most of them will have intrinsic ellipticities of $\sim 0.55$. In this case almost no intrinsically spherical systems are found and a rather large percentage of highly flattened systems is predicted. In the case of a purely prolate distribution ($T = 1.0$), the average intrinsic ellipticity is also $\sim 0.55$. However, slightly more clusters with small elongations are present than in the oblate case, and less with high elongations. When only intermediate triaxial clusters (having $T = 0.5$) are considered, the profile is intermediate between those for oblate and prolate shapes. If we try to reconstruct the profile over $\epsilon_{true}$ from these constant-$T$ distributions, the $T = 0.5$ and $T = 1.0$ distributions perform best. The purely oblate population produces too many spherical clusters. This result is in agreement with the analysis of Salvador-Solé & Solanes (1993), who find the projected elongations of clusters to be consistent with a prolate distribution with intrinsic axial ratios Gaussian-distributed with mean $\sim 0.5$ and standard deviation $\sim 0.15$. These elongations are mainly produced by the tidal interactions of massive enough nearby clusters.

Because the density parameter $\Omega$ will most certainly have an influence on the shapes of clusters, we use N-body simulations, which include a recipe for galaxy formation and merging. A CDM-spectrum and a density parameter $\Omega = 1.0$ are used. The biasing parameter found for these simulations is $b = 2.0$. The simulations form a complete catalogue that is expected to give a fair representation of all real clusters. The projected galaxy distributions follows roughly the projected Hubble-profile, eq. (10). 'Observing' these clusters in the same way as the real clusters produces too large elongations as compared to the observations. For $\Omega = 0.2$ a similar catalogue is being prepared, though not ready yet. We have picked out nine of the $\Omega = 0.2$ clusters (both massive and less massive ones) that are simulated in the same way as their $\Omega = 1.0$ counterparts, using the same random number seeds. A biasing parameter $b = 1.0$ is adopted here. The $\Omega = 0.2$ clusters are, on average, less elongated than their $\Omega = 1.0$ counterparts. The cases where the $\Omega = 0.2$ cluster is more elongated turn out to be due to merging, projection of two clusters onto each other, or because the $\Omega = 0.2$ cluster actually is a double cluster. The preliminary results of the N-body simulations suggest that $\Omega = 0.2$ is more consistent with the data than for $\Omega = 1.0$. A more quantitative analysis will be possible when the whole $\Omega = 0.2$ catalogue will be finished. The results of the simulations are consistent with the results of Evrard (1993) and Evrard et al. (1993), who use simulations of the X-rays tracing the cluster potential. They also found low-$\Omega$ clusters to be more nearly spherical and also more centrally condensed than $\Omega = 1.0$ clusters. However, Evrard (1993) concludes that their $\Omega = 1.0$ simulations give a better description of the observations than do the $\Omega = 0.2$ simulations, although they have only 4 clusters to compare with. A follow-up paper with a sample of 50-60 *EINSTEIN*-clusters is in progress (Mohr et al. 1994).

As the $\Omega = 0.2$ catalogue is not complete yet, we use the $\Omega = 1.0$ catalogue to calculate the intrinsic shapes of the model clusters. They turn out to be mainly nearly prolate, with more than 50% having a triaxiality parameter larger than 0.8. Almost no (nearly) oblate clusters are detected. This is consistent with what is expected from the distribution of true projected cluster shapes.

**Acknowledgments**

The authors would like to thank Tim de Zeeuw and Simon White for useful discussions and a careful reading of the manuscript. Marijn Franx is acknowledged for useful discussions.

The comments of the referee, August Evrard, were very useful and helped to improve this paper.